# Bayesian inference in band excitation Scanning Probe Microscopy for optimal dynamic model selection in imaging

Rama K. Vasudevan,[1,1] Kyle P. Kelley,[1] Eugene Eliseev,[2] Stephen Jesse,[1] Hiroshi Funakubo,[3] Anna Morozovska,[4] and Sergei V. Kalinin[1,2]

[1] The Center for Nanophase Materials Sciences, Oak Ridge National Laboratory,
Oak Ridge, TN 37831

[2] Institute for Problems of Materials Science, National Academy of Sciences of Ukraine,
Krjijanovskogo 3, 03142 Kyiv, Ukraine

[3] Department of Materials Science and Engineering, Tokyo Institute of Technology, Yokohama, 226-8502, Japan

[4] Institute of Physics, National Academy of Sciences of Ukraine,
46, pr. Nauky, 03028 Kyiv, Ukraine

**Abstract**

The universal tendency in scanning probe microscopy (SPM) over the last two decades is to transition from simple 2D imaging to complex detection and spectroscopic imaging modes. The emergence of complex SPM engines brings forth the challenge of reliable data interpretation, i.e. conversion from detected signal to descriptors specific to tip-surface interactions and subsequently to materials properties. Here, we implemented a Bayesian inference approach for the analysis of the image formation mechanisms in band excitation (BE) SPM. Compared to the point estimates in classical functional fit approaches, Bayesian inference allows for the incorporation of extant knowledge of materials and probe behavior in the form of corresponding prior distribution and return the information on the material functionality in the form of readily interpretable posterior distributions. We note that in application of Bayesian methods, special care should be made for proper setting on the problem as model selection vs. establishing practical parameter equivalence. We further explore the non-linear mechanical behaviors at topological defects in a classical ferroelectric material, PbTiO$_3$. We observe the non-trivial evolution of Duffing resonance frequency and the nonlinearity of the sample surface, suggesting the presence of the hidden elements of domain structure. These observations suggest that the spectrum of anomalous behaviors at the ferroelectric domain walls can be significantly broader than previously believed and can extend to non-conventional mechanical properties in addition to static and microwave conductance.

---
[1] vasudevanrk@ornl.gov
[2] sergei2@ornl.gov

Scanning probe microscopy techniques have emerged as one of the primary tools for exploring materials and devices on the nanometer, molecular, and atomic levels.[1-7] Examples of structural imaging enabled by SPM include surfaces of metals,[8-10] oxides,[11, 12] semiconductors,[13-15] polymers,[16-18] and complex biological systems.[19-23] Equally broad is the spectrum of SPM applications for functional imaging, providing real space data on the electrical,[24-26] magnetic,[27-30] mechanical,[31-33] ferroelectric,[34-36] optoelectronic,[37, 38] and other functional properties in broad range of materials systems.

The universal tendency in SPM imaging modes over the last two decades is to transition from simple 2D imaging to complex detection and spectroscopic imaging modes. The latter include the gamut of force-distance spectroscopies in atomic force microscopy (AFM),[39] current spectroscopies in scanning tunneling microscopy (STM),[40, 41] and complex time and voltage spectroscopies in piezoresponse force microscopy (PFM)[42-44] and electrochemical strain microscopies.[45] These spectroscopies define the parameter space sampled at each spatial pixel during a SPM experiment, where the detection modes define the nature of the signal measured. Their evolution is exemplified by the transition from simple lock-in and phase lock loop based detection schemes that yield a scalar response in each point of the parameter space, to more complex band excitation,[46, 47] intermodulation,[48] and G-Mode SPMs.[49-51]

The proliferation of complex SPM engines brings forth the challenge of the reliable data interpretation, i.e. conversion from detected signal to descriptors specific to tip-surface interactions and subsequently to materials properties. In the band excitation (BE) family of SPM techniques, the analysis is traditionally based on the simple harmonic oscillator (SHO) fit of response amplitude and phase vs. frequency data.[52] This fitting yields resonant frequency, amplitude, phase difference between drive and response, and quality factor that define the response and energy loss at the tip-surface junction. The introduction of BE allowed SPM to avoid the under determinedness of the tip dynamics inevitable in classical single frequency methods and enabled quantitative cross-talk free[42] imaging. In turn, these parameters can be linked via contact mechanic models to the materials properties.

However, data analysis in BE until now relied exclusively on the SHO model. While examples of more complex dynamic behaviors are frequently observed, the analysis in terms of more complex models have been impractical. The reason for this is that while many non-linear models allow for approximate solutions of the frequency response, the model selection is an open challenge. Furthermore, functional fits of individual responses give rise to large uncertainties in fitting parameters, resulting in extremely high noise in output maps. Most importantly, the functional behaviors of the response can differ across the sample surface, precluding the use of single *ad hoc*-chosen model for analysis of hyperspectral data.

Here, we introduce an approach for quantification of basic BE data using Bayesian regression, allowing both for model selection and determination of model parameters. As natural for Bayesian methods, this allows for incorporation of the prior knowledge of the microscope and materials behavior. This approach yields the local point estimates of required responses, along with the posterior probabilities for parameter values. Here, we demonstrate this approach for comparison of SHO and Duffing model for low values of simulated nonlinearity, but note that a similar approach can be implemented using more complex parametric models or numerical solvers. We further highlight its application to real experimental data captured in a ferroelectric system.

The Bayesian approach for inference is based on the concepts of prior and posterior probabilities linked via Bayes formula:[53, 54]

$$p(\theta_i|D) = \frac{p(D|\theta_i)p(\theta_i)}{p(D)} \qquad (1)$$

Here $D$ represents the data obtained during the experiment, $p(D|\theta_i)$ represents the likelihood that this data can be generated by the theory, e.g. given a choice of model $i$, and model parameters $\theta$. $p(\theta_i)$ is the prior, i.e the probability function for the model and model parameters. Finally, $p(D)$ is the denominator that defines the total space of possible outcomes.

Note that despite the intrinsic elegance of Bayes approach and its transparent scientific meaning, its adoption by many scientific communities has been exceedingly slow. This happens for two primary reasons. First, evaluation of denominator in Eq. (1) requires very high dimensional integrals and become feasible for experimentally relevant distributions only in the last several years. Secondly, the choice of priors can be a contentious issue, unsurprisingly given that the vast majority of application to date has been concentrated in statistical, medical, and sociological communities where priors can be difficult to define. Interestingly, in the physics field, domain knowledge is sufficiently developed to provide meaningful priors, making the analysis via Eq. (1) well suited to typical scientific workflows. Notably, the unique strength of the Bayesian approach is that model selection can be incorporated as a part of the inference process. In this case, models can be drawn from a list of possible models with certain probability, and posterior distribution will update this probability to redefine model selection.

Here, we develop this framework for BE PFM, the technique for which data analysis pivots on the appropriate fit model selection. The data was acquired using an Oxford Instruments Cypher AFM integrated with LabView framework (see experimental section). As a model material system, we have chosen $PbTiO_3$ films with a dense a-c domain structure, providing multiple topological defects and hence offering the potential for uncovering illusive physics present in the form of nonlinear responses.

The representative surface morphology is presented in Figure 1a showing topographical features as large as 30 nm. The corresponding BE PFM resonant frequency, phase, amplitude, and quality factor are shown in Figure 1(b-e), respectively and demonstrate rich domain structure formed by the large-scale c+ and c- domain and near surface in-plane a-domain forming a clearly visible mesh-like structure. Note, the BE PFM images presented here were assembled using a traditional SHO fits. In the analysis presented here, we developed a Bayesian inference framework to further understand BE generated data.

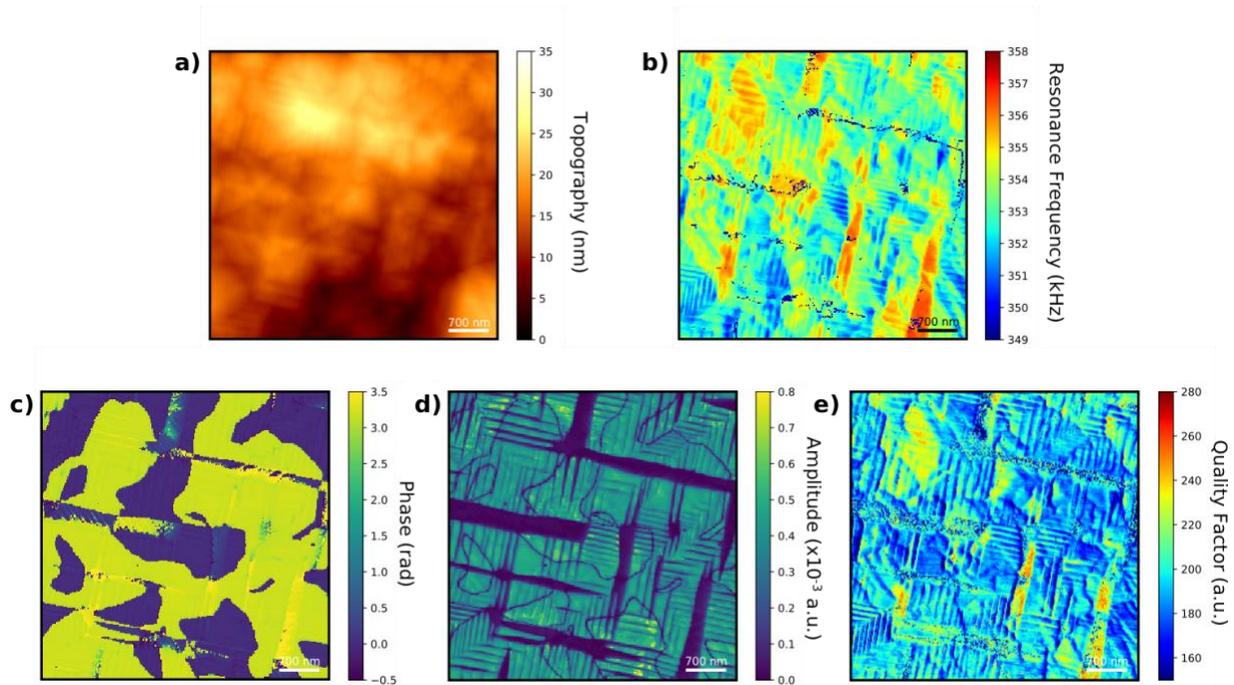

**Figure 1.** a) Surface topography of 700 nm thick $PbTiO_3$ thin film and corresponding BE PFM b) resonance frequency, c) phase, d) amplitude, and e) quality factor derived from SHO fits.

To perform the Bayesian fit of BE data, we develop Bayesian framework for models with known analytical (or approximate) solutions. Here, we consider two primary models, the simple harmonic oscillator (SHO) and Duffing model. Notably, the SHO model is a special case of the Duffing model; hence separation between the two is not a classification or model selection problem. Rather, it represents a practically equivalent case where the determination of a chosen parameter (specifically, nonlinearity) tends to zero.

The SHO model is defined by the equation $m\ddot{u} + \gamma\dot{u} + ku = f\sin(\omega t)$ where $m$ is the oscillator mass, $\gamma$ is damping coefficient, $k$ is spring constant, and $f\sin(\omega t)$ is the periodic driving force. The SHO model has an analytical solution, where the frequency dependence of response is given by

$$R(\omega) = \frac{f\omega^2 e^{i\varphi}}{\omega^2 - \frac{i\omega\omega_r}{Q} - \omega_r^2} \qquad (2)$$

where ω is the frequency, $\omega_r$ is the resonant frequency, Q is the quality factor, and φ is the phase. Previously, all analysis of the BE data was based on the direct, least-squares fit of Eq. (2) to the experimental data,[42] with the fit parameters plotted as the maps as shown in Figure 1. Note, in this manuscript we explicitly do not treat the phase of the response and will refer only to the amplitude.

In comparison, the Duffing model[55] allows for the presence of cubic non-linearity in the tip-surface interactions, and is given by

$$\mu\frac{d^2u}{dt^2} + c\frac{du}{dt} + ku + \lambda u^3 = f\sin(\omega t), \qquad (3)$$

where we denote the displacement from the equilibrium state as "$u$", effective "mass" of the oscillator as "$\mu$" ($\mu > 0$), "c" is the damping coefficient, while "$k$" and "$\lambda$" are for linear and nonlinear stiffness coefficients respectively.

Depending on the sign and value of the cubic term, the Duffing model can rise to multiple regimes including chaotic oscillations, jumps, etc.[56, 57] Here, we consider the case of small cubic term, where the approximate parametric solution for amplitude-frequency curve can be found assuming a quasi-harmonic approximation, $u \approx R\sin(\omega t + \varphi)$, as

$$R^2\left(4(\widetilde{\omega} - 1)^2 + \tilde{c}^2 + 3(1 - \widetilde{\omega})\tilde{\lambda}R^2 + \frac{9}{16}\tilde{\lambda}^2 R^4\right) = \tilde{f}^2, \qquad (4)$$

where we introduced the following dimensionless parameters

$$t = \tilde{t}\sqrt{\frac{\mu}{k}}, \qquad \widetilde{\omega} = \omega\sqrt{\frac{\mu}{k}}, \qquad \tilde{c} = \frac{c}{\sqrt{k\mu}}, \qquad \tilde{\lambda} = \frac{\lambda}{k}, \qquad \tilde{f} = \frac{f}{k}. \qquad (4c)$$

Derivation of Eq. (4) is given in the Supplementary Materials, Appendix A. The comparison of the approximate Eq. (4) and the amplitude of the first harmonics of the numerical solution of Eq. (3) is shown in Figure 2a for several values of driving force amplitude $\tilde{f}$.

To deconvolute the experimental results, we derived from the parametric Eq. (4) approximate analytical dependence for the amplitude-frequency curve

$$R \approx \sqrt{2}\tilde{f}\left(4(\widetilde{\omega}-1)^2 + \tilde{c}^2 + \sqrt{(4(\widetilde{\omega}-1)^2 + \tilde{c}^2)^2 + 3(1-\widetilde{\omega})\tilde{\lambda}\tilde{f}^2}\right)^{-\frac{1}{2}} \qquad (5)$$

The approximate solution Eq. (5) is compared with numerical solution of Eq. (4) in Figure 2b, showing excellent agreement between the two.

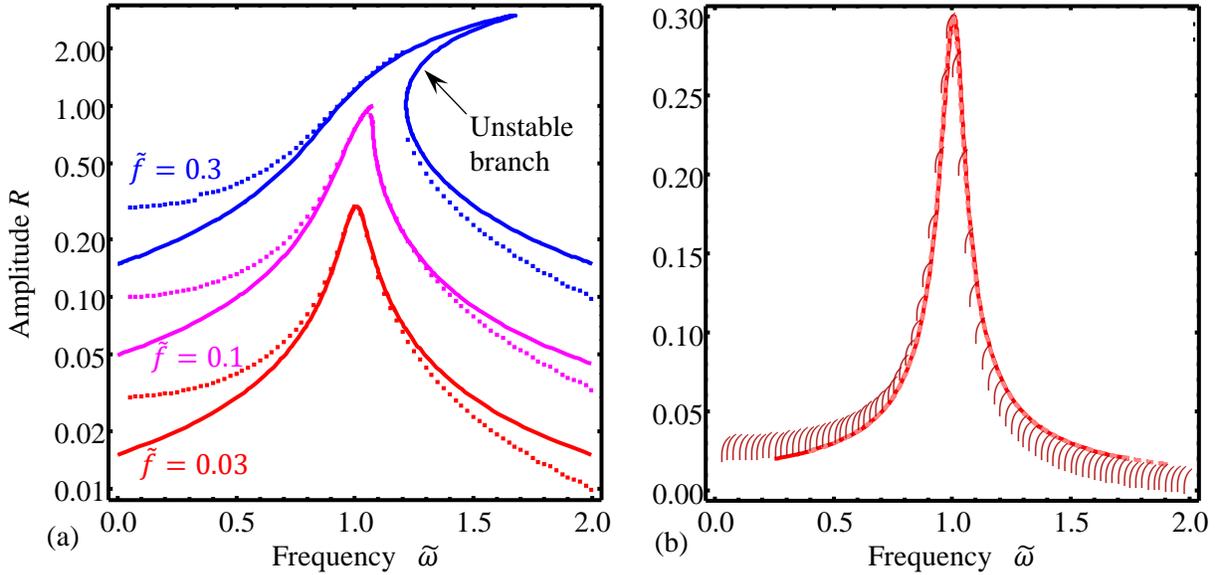

**Figure 2. (a)** Dependence of the oscillation amplitude on the driving force frequency for the different amplitude of force, $\tilde{f} = 0.03, 0.1, 0.3$ (red, magenta and blue curves respectively). Dotted curves are numerical solution of Eq. (3), solid curves are the graphical solution of Eq.(4b). **(b)** Dependence of the oscillation amplitude on the driving force frequency for the different amplitude of force, $\tilde{f} = 0.03$. Other parameters: $\tilde{c} = 0.1$, $\tilde{\lambda} = 0.2$. Dots are numerical solutions of Eq.(4b), solid and dashed curves are Eq. (5).

To perform Bayesian inference fit, we have implemented Bayesian regression utilizing the PyMC3 library available in Python for both the SHO and Duffing oscillator equations.[58] Note that the SHO is a subset of the Duffing oscillator. The difference arising purely from the nonlinearity term, such that, in the limit $\lambda \to 0$, the Duffing model parameters converge to the SHO. Hence, differentiating Duffing vs. SHO behavior becomes the problem of finding the posterior probability density $\lambda$ and defining the region of practical equivalence (ROPE)[53] of $\lambda$ to 0.

Here, we utilized the Metropolis-Hastings (MH) algorithm to sample the distributions of model parameters of both models with the assumption that the measured response data can be modeled as noisy measurements of the form $y_i = R_i + \eta$ where the error η comes from a normal distribution with a variance of $\sigma^2$, i.e. $\eta \sim N(\mu, \sigma^2)$. Therefore, in addition to the model parameters, the data variance is also estimated. The priors for the two models are given in Table 1.

**Table 1: Priors used in sampling**

| | SHO Equation Priors | | Duffing Equation priors |
|---|---|---|---|
| **Amplitude** | $f \sim Uniform(0,1)$ | **Reduced Forcing** | $\tilde{f} \sim Uniform(0,0.1)$ |
| **Resonance** | $\omega_r \sim Uniform(\min(\omega), \max(\omega))$ | **Reduced Nonlinearity** | $\tilde{\lambda} \sim Normal(0,0.1)$ |
| **Quality Factor** | $Q \sim Uniform(0,300)$ | **Reduced Damping** | $\tilde{c} \sim HalfNormal(0,0.1)$ |
| **Data Variance** | $\eta \sim Gamma \begin{pmatrix} \alpha = 0.001, \\ \beta = 0.001 \end{pmatrix}$ | **Data Variance** | $\eta \sim Gamma \begin{pmatrix} \alpha = 0.001, \\ \beta = 0.001 \end{pmatrix}$ |

Note: Gamma distribution takes the form $f(x|\alpha,\beta) = \frac{\beta^\alpha x^{\alpha-1} e^{-\beta x}}{\Gamma(\alpha)}$

To illustrate the principle and performance of Bayesian regression, we perform analysis on simulated data varying the non-linearity and noise levels with parameters of both models chosen to be similar (i.e., to make distinguishability difficult). Shown in Figure 3(a,b) are examples of the SHO and Duffing responses for two different noise levels. The corresponding fits, taken by utilizing the mean and standard deviations of the model parameters, are shown as solid and dashed lines, respectively in Figure 3(c,d). A Jupyter notebook that allows exploring the synthetic data analysis is provided.

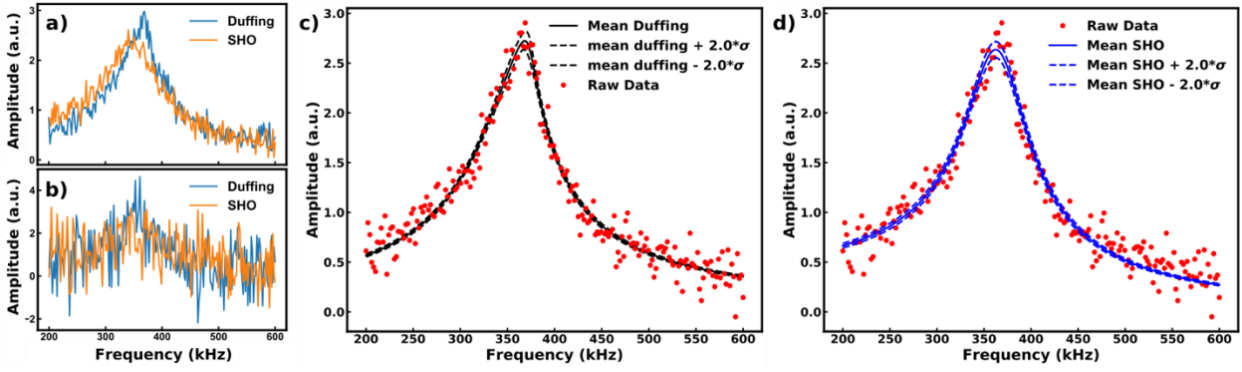

**Figure 3**. **Bayesian Regression on synthetic data for two noise levels.** **(a)** Modeled SHO Curve (orange) with parameters $[f, \omega_r, Q, \sigma^2] = [0.58, 352 \times 10^3, 4, 0.15]$, and Modeled Duffing Oscillator response with parameters $[\tilde{f}, \tilde{c}, \tilde{\lambda}, \sigma^2] = [0.5, 0.2, 0.06, 0.15]$ (blue line). **(b)** SHO (orange) and Duffing (blue) responses for same parameters except for higher noise ($\sigma^2 = 0.9$). We performed sampling to determine parameters of the SHO and Duffing model based on the synthetic Duffing curve in (a), with results for both the SHO model and Duffing model Bayesian Regression plotted in **(c)** and **(d)**, respectively. These 'fits' are derived from means and standard deviations calculated from the full posteriors of the parameters of the respective models. The derived estimates are shown in Supplementary Table 1. Unsurprisingly, the parameters are well recovered for the low noise case, but the variance of the estimates increases for the higher noise case (Supplementary Materials Table 1).

In contrast to standard least squares curve fitting, which provides only the point estimate of mean values and covariance, the Bayesian method allows us to obtain the full posterior densities. For insight into how this process operates, we created a synthetic dataset using the Duffing equation where we varied both the degree of nonlinearity ($\tilde{\lambda}$) as well as the amount of noise. We kept the parameters $\tilde{f} = 0.5, \tilde{c} = 0.2$ and varied $\tilde{\lambda}$ linearly on the interval [0, 0.05], and varied the noise on the interval [0,1] to produce a 2D response map. Performing MH sampling results in posteriors that can be plotted as a function of noise (or lambda) in 2D form, as shown in Figure 4. It is clearly seen that the increasing noise causes spreading of the posterior densities (higher variance), as would be expected. Interestingly, the linearly increasing data variance is almost perfectly modeled in these maps.

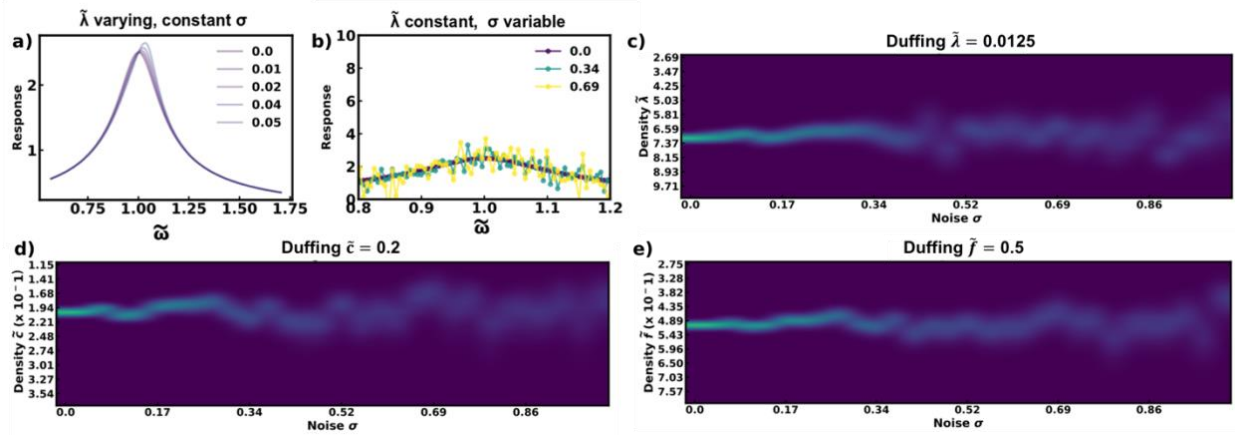

**Figure 4. 2D Posterior Densities for both Duffing Model for fixed nonlinearity and increasing noise. (a)** Simulated data for fixed noise, but varying nonlinearity. (b) Simulated data for fixed nonlinearity, but varying noise. (c-e) 2D posterior densities are shown for the four estimated parameters for the Duffing model (Ground truth is indicated in the title for (c-e). The meandering of the density and increase in variance are obvious.

As a comparison, we further compute the SHO model parameters for the same 2D synthetic dataset. Provided the two sets of posterior densities (one for each model), we may then estimate the probability of the models using the widely-applicable information criterion (WAIC)[59] for both models, which is defined by Gelman et al.[60] to be a generalized version of the Akaike information criterion, and crucially can be numerically calculated without knowledge of the true underlying distribution. The WAIC starts with the log predictive density (lppd) with a correction factor for the effective number of parameters of the model. The lppd is calculated as:

$$lppd = \sum_{i=1}^{n} \log\left(\frac{1}{S}\sum_{s=1}^{S} p(y_i|\theta^s)\right) \tag{6}$$

where there are $S$ simulation draws. The calculation of the effective number of parameters $pWAIC$ is given by summing the variance of the log likelihood, $\log p(y_i|\theta)$ for each of the $n$ datapoints available.

$$p_{WAIC} = \sum_{i=1}^{n} var_{post}(\log p(y_i|\theta)) \tag{7}$$

The difference between these two provides the WAIC. It should be emphasized that such metrics require the use of the traces acquired during the sampling and not just the point estimates to calculate (6) and (7). Once the WAIC is calculated for both models the probability of the model p(M) is recovered via a Bayesian model averaging approach that utilizes pseudo-BMA using AIC-type weighting as discussed in Ref. [61].

The corresponding *p(M)* map for the 2D synthetic dataset is shown in Figure 5. The Duffing model is preferred for all cases (probability *p*>0.5), but the distinguishability becomes more difficult in higher noise settings. As expected, the higher the nonlinearity, the more the Duffing model is preferred for a given noise level, although the effect is rather weak. This is probably due to the limited nonlinearity range explored in this simulation (analytical approximations will also break down beyond this level of $\tilde{\lambda}$).

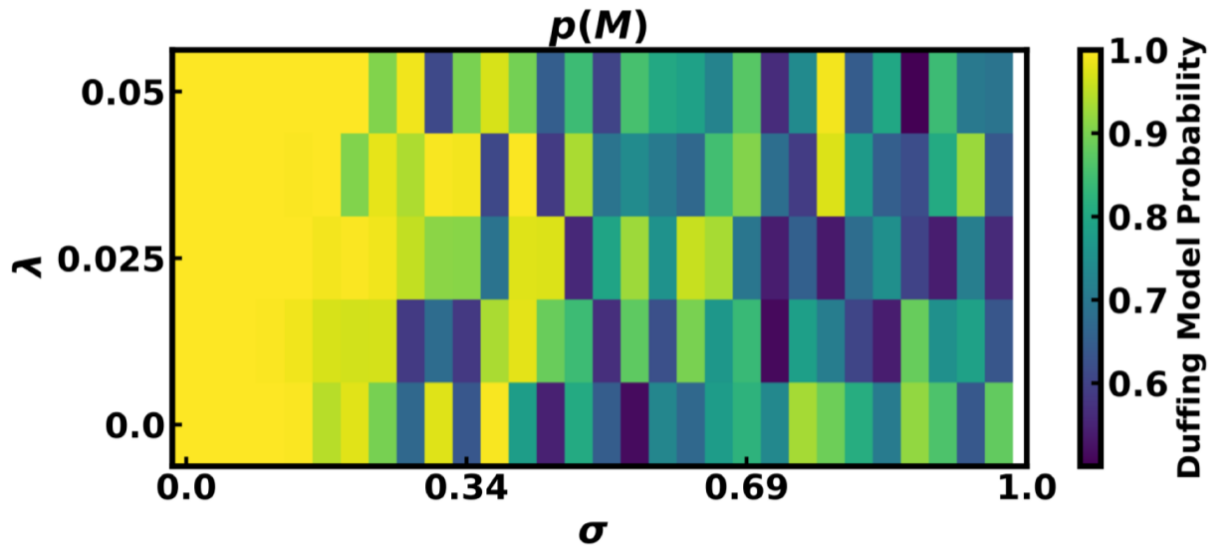

**Figure 5.** Model selection as a function of varying nonlinearity ($\tilde{\lambda}$) and noise level ($\sigma$). The Duffing model is preferred in all cases. Note that the ground truth model is Duffing.

**Experimental Results**

We now turn to the use of this methodology to explore the experimental results presented in Figure 1. Since full MCMC sampling is computationally expensive, we restrict ourselves to a small portion of the data cube, highlighted by the boxed region in Figure 6(a). The close-up of this region is shown in Figure 6(b) in the BE PFM amplitude image (fit using traditional SHO model least squares method). The means of the posteriors of the four parameters of the Duffing model for this region are shown below in Figure 6(c-f). Firstly, the image of the linear stiffness parameter appears to show details that are not present within the original SHO fit amplitude image, i.e. there is a clear qualitative difference. Next, the nonlinearity ($\tilde{\lambda}$) map indicates nonzero values for most of the map, suggesting that nonlinearities of some sort do exist within the system. The BE-PFM Duffing amplitude map in Figure 6(d) shows characteristically similar features to Figure 6(b). It should be noted that these cannot be compared directly because of the reduced units used; nonetheless, qualitatively they look very similar. Finally, the estimate of the variance shows that the areas with higher signal (away from domain walls) actually have higher variance. This can be rationalized by observing that when signals are amplified by the quality factor of the cantilever, this does not cancel out all noise – indeed, some of the noise is also amplified, leading to this somewhat counterintuitive result that domain walls display lower variance than the actual domains themselves.

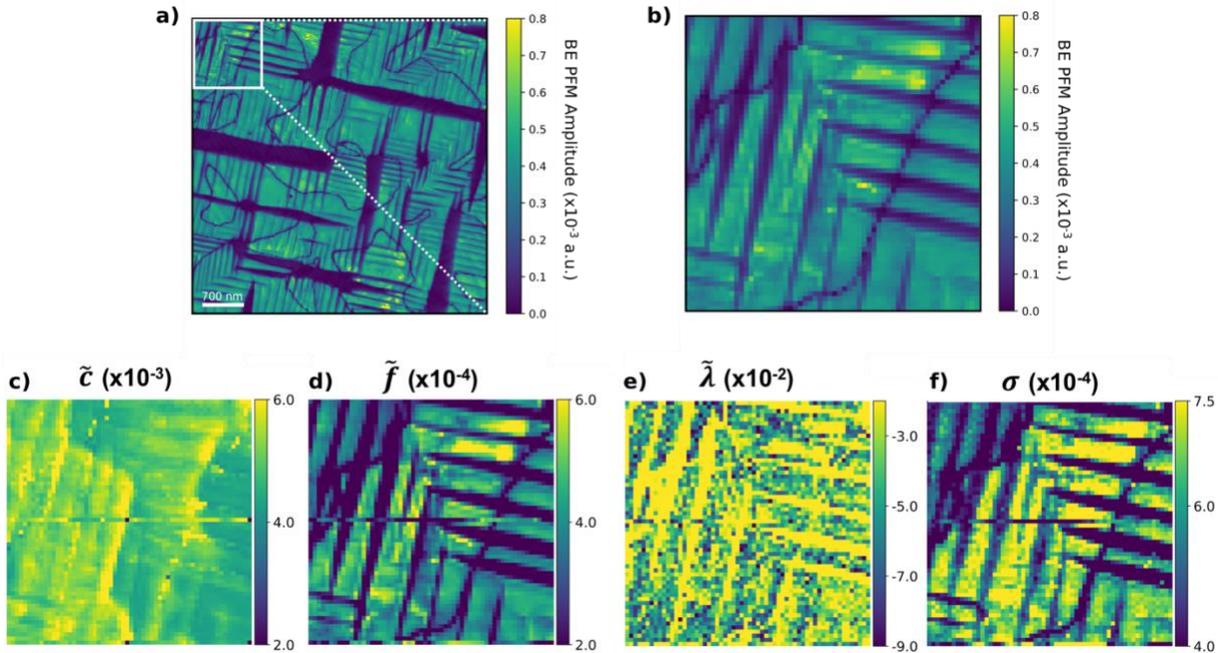

**Figure 6**. BE-PFM data with Duffing oscillator Bayesian regression performed on a subset of the data. The fitting maps are shown below (only means). (a) BE-PFM Amplitude image of a 5x5um area of the sample. (b) Subset taken from (a). (c-f) Point estimates (mean) of the posterior probabilities for the four parameters of the Duffing model for this data subset.

Since the model was sampled with MCMC, we have access to the full traces and can plot posterior densities as in the simulated case. This provides more guidance on where such model point estimates are deemed reliable, and where the variance is very large (i.e., unreliable). Shown in Figure 7(a) is the map of the nonlinearity parameter (repeat of Figure 6(e)). A line profile is taken through the 15$_{th}$ row as indicated by the dashed line. We plot the 2D posterior density with the x-axis being along the line profile direction, and the y-axis indicating the weight of the posterior (i.e., essentially a histogram). It is observed that the nonlinearity parameter varies along this line, but more importantly, it is more strongly compressed in some areas (tighter distributions) and more spread out over others, where more uncertainty exists. The inferred variance in Figure 7(c) also oscillates, agreeing with the domain structure as described earlier.

**Model Selection**

To observe where the SHO oscillator model may be insufficient to describe the measured response, we explored the locations where the nonlinearity $|\tilde{\lambda}| > 0.07$. These points are mapped onto the BE-PFM amplitude image in Figure 7(d). We then computed the probability of the SHO model compared and the Duffing model for each red pixel in Figure 7(d) and found that the Duffing model was preferred in all cases with weights generally over 0.99. A few selected locations with raw data, and Duffing and SHO mean estimates are shown in Figure 7(e-g). Although it is somewhat difficult to make sense of the locations of these pixels, many appear on the left side of the domain wall, which may be due to the preferential depinning of the walls and known asymmetry of ferroelastic wall motion in these structures.[62]

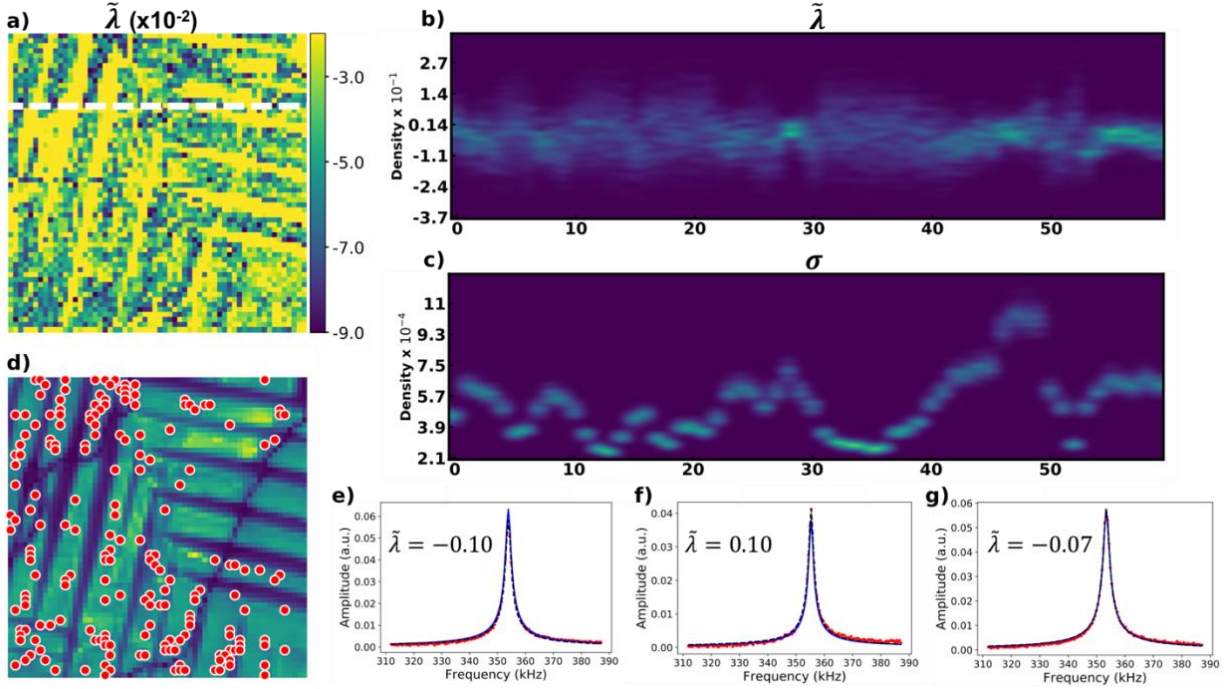

**Figure 7.** Profiles and model selection. (a) nonlinearity map. (b) 2D posterior density along the line in (a) for lambda. (c) 2D posterior density along the line in (a), for variance. (d) Location of all pixels where |lambda|>0.07. (e-g) Raw data at randomly selected red pixel locations in (d) with Duffing model estimate (black) and SHO Model estimate (blue).

To gain insight into possible origins of the non-linear mechanical properties in the ferroelectric film, we consider simplified model that couples one-component ferroelectric polarization *P* with elastic strain *u*:

$$\sigma \cong \frac{1}{s}u - \frac{d}{s}P - \frac{Q}{s}P^2, \tag{8a}$$

$$\mu\frac{\partial^2 P}{\partial t^2} + \Gamma\frac{\partial P}{\partial t} + 2(a - Q\sigma)P + bP^3 + wP^5 - g\frac{\partial^2 P}{\partial x^2} = d\sigma + E(x,t), \tag{8b}$$

Here, the Eq. (8a) is Hooke's law relating the stress $\sigma$ with elastic strain *u*, where *d* is an effective piezoelectric constant, *Q* is an electrostriction coefficient and *s* is an elastic constant. Eq.(8b) is a one-component and one-dimensional nonlinear time-dependent LGD equation for *P*, where the kinetic coefficients $\mu$ and $\Gamma$ are responsible for the polarization stiffness (e.g. domain walls elasticity) and viscosity (Khalatnikov relaxation). (see Supplementary Materials, Appendix B for details). Note that we consider uncharged domain walls, which do not induce depolarization and local electric fields, so that $E(x,t)$ coincides with applied external field $E_{ext}(t)$. We further assume that the material is deep in a ferroelectric phase, $P = P_S + \delta P$, where the spontaneous polarization $P_S$ is enough high, and so piezoelectric reaction dominates over electrostriction response, $\left|\frac{d}{s}P_S\right| \gg \left|\frac{Q}{s}P_S^2\right|$. Since the normal stresses are absent at the elastically free surface, the piezo-strain is $u \approx dP$ in the region of BE-PFM response formation. We further assume that electrostriction contribution is negligibly small in comparison with effective piezoelectric coupling and soft mode nonlinearity. Also, we assume that the spatial dispersion has a characteristic period $k_0$ in a Fourier space, and so we can estimate that $\frac{\partial^2 P}{\partial x^2} \approx -k_0^2 P$. With these simplifications Eq.(8b) acquires the form:

$$\mu \frac{\partial^2 u}{\partial t^2} + \Gamma \frac{\partial u}{\partial t} + ku + \lambda u^3 = f_{ext}(t), \qquad (9)$$

where $k(T) = 2a(T) - \frac{d^2}{s} + gk_0^2$, and $\lambda = \left(b + 2\frac{Q^2}{s}\right)\frac{1}{d^2}$. The detailed derivation is provided in Supplementary Materials. Note that Eq. (9) is a material equation, it can model the internal strain, but not the measured BE-PFM response, that is a convolution of the strain tensor with corresponding transfer function of a tip-surface junction.

Let us make some estimates of the coefficient $\lambda$ for the ferroelectric solid solution PbZr$_x$Ti$_{1-x}$O$_3$ (PZT, x<0.5), which can be the first order for x<0.2 and the second order for 0.3<x<0.5 at room temperature. Using its parameters from Refs.[63-65], we obtain the ranges for $Q \approx (0.01 - 0.05)$ m4/C2, $s \approx 1.5 \cdot 10$-11 m2/N, $a \approx (10$8 $- 10$9$)$ m2N/C2, $b = \pm (0.05 - 5)10$7m6N/C4, $g \approx 10$-10C-2m4N, $k_0^2 \approx (1 - 10) \cdot 10^{18}$ m-2, and $d \approx (2.3 - 5) \cdot 10$-2m2/C [66], which give $k = \pm(0.5 - 5) \cdot 10^9$m2N/C2 and $\lambda = \pm(0.5 - 50) \cdot 10^9$ m2N/C2. Since the sum $\left(b + 2\frac{Q^2}{s}\right)$ can be negative, almost zero, or positive for the ferroelectrics with the first order phase transition in the bulk (i.e. when b<0), as well as change its sign due to the electrostriction coupling (adding positive term $2\frac{Q^2}{s}$), the dimensionless nonlinearity $\tilde{\lambda} = \left|\frac{\lambda}{k}\right| = (0.01 - 100)$ can vary in a wide range, and hence can provide explanations for observed behaviors.

Even the simplified 1D analysis suggest that ferroelectric behavior can be an origin of mechanical non-linearity. In the vicinity of the domain walls, the wall deformation can be additional source of non-linear behavior; however, the numerical analysis in this case will require adaptation of the full 2- or 3D phase field codes and is outside of the scope of this work.

To summarize, here we implemented the Bayesian inference approach for the analysis of the image formation mechanisms in scanning probe microscopy. Compared to the point estimates in classical functional fit approaches, Bayesian inference allows to incorporate prior knowledge of materials and probe behavior in the form of corresponding prior distribution and return the information on the material functionality in the form of readily interpretable posterior distributions. We note that in the application of Bayesian methods, special care should be taken for proper setting of the problem as model selection vs. determination of equivalence. The former problem corresponds to determination of model probability via well-established numerical criteria whereas the second necessitates operator-made definitions.

Using these the Bayesian inference approach allows exploration of non-linear behavior in classical ferroelectric $PbTiO_3$ material. We observe the non-trivial evolution of Duffing resonance frequency and the nonlinearity of the sample surface, suggesting the presence of the hidden elements of domain structure. These observations suggest that the spectrum of anomalous behaviors at the ferroelectric domain walls can be significantly broader than believed previously and can extend to non-conventional mechanical properties in addition to static and microwave conductance.[67, 68]

In the future, we aim to extend this approach to inferential analysis using direct numerical solutions. While too slow for practical use now, the incorporation of tabulated and machine-learning interpolations are potential venues for development.

**Experimental Methods**

As a material system, we have chosen a 700 nm thick $PbTiO_3$ thin film grown by chemical vapor deposition on (001) $KTaO_3$ substrates with a $SrRuO_3$ conducting buffer layer, as reported by H. Morioka et al.[69, 70] The PFM was performed using an Oxford Instrument Asylum Research Cypher microscope with a National Instruments DAQ card and chassis, and operated with a LabView framework. All experiments were performed using Budget Sensor Multi75E-G Cr/Pt coated AFM probes (~3 N/m). All band excitation data was acquired with an AC excitation voltage of 2V.

**Acknowledgements**

This research was conducted at the Center for Nanophase Materials Sciences, which also provided support (R. K. V, S. J., S. V. K.) and is a US DOE Office of Science User Facility. The PFM portion of this work was supported by the U.S. Department of Energy, Office of Science, Materials Sciences and Engineering Division (K. K.). This research used resources of the Compute and Data Environment for Science (CADES) at the Oak Ridge National Laboratory, which is supported by the Office of Science of the U.S Department of Energy under Contract No. DE-AC05-00OR22725.

**Data Availability**

The data that support the findings of this study are available from the corresponding author upon reasonable request.

# Supplementary Materials to

"Bayesian inference in band excitation Scanning Probe Microscopy for optimal dynamic model selection in imaging"

## Appendix A. Duffing oscillator model

Let us consider the oscillations of some system in the nonlinear potential:

$$\mu \frac{d^2 u}{dt^2} + c \frac{du}{dt} + ku + \lambda u^3 = f \sin(\omega t) \tag{A.1}$$

Here we denote the displacement from the equilibrium state as "$u$", effective "mass" of the oscillator as "$\mu$" ($\mu > 0$), "$c$" is the damping coefficient, while "$k$" and "$\lambda$" are for linear and nonlinear stiffness coefficients respectively.

Dimensionless equation is

$$\frac{d^2 u}{d\tilde{t}^2} + \tilde{c} \frac{du}{d\tilde{t}} + u + \tilde{\lambda} u^3 = \tilde{f} \sin(\tilde{\omega} \tilde{t}) \tag{A.2a}$$

Where we introduced the following dimensionless parameters

$$t = \tilde{t}\sqrt{\frac{\mu}{k}}, \quad \tilde{\omega} = \omega\sqrt{\frac{\mu}{k}}, \quad \tilde{c} = \frac{c}{\sqrt{k\mu}}, \quad \tilde{\lambda} = \frac{\lambda}{k}, \quad \tilde{f} = \frac{f}{k}. \tag{A.2b}$$

The approximate solution for the small values of driving force amplitude and nonlinearity stiffness coefficient could be obtained via so called "two variable expansion method" [i], giving the relation between the amplitude of periodic solution,

$$u \approx R \sin(\omega t + \varphi)$$

and the driving force frequency $\omega$:

$$\tilde{\omega} = 1 + \frac{3}{8}\tilde{\lambda}R^2 \pm \frac{1}{2}\sqrt{\left(\frac{\tilde{f}}{R}\right)^2 - \tilde{c}^2} \tag{A.3}$$

The comparison of the approximate expression (A.3) and the amplitude of the first harmonics of the numerical solution of Eq.(A.2a) is shown in Fig. S1 for the several values of driving force amplitude $\tilde{f}$.

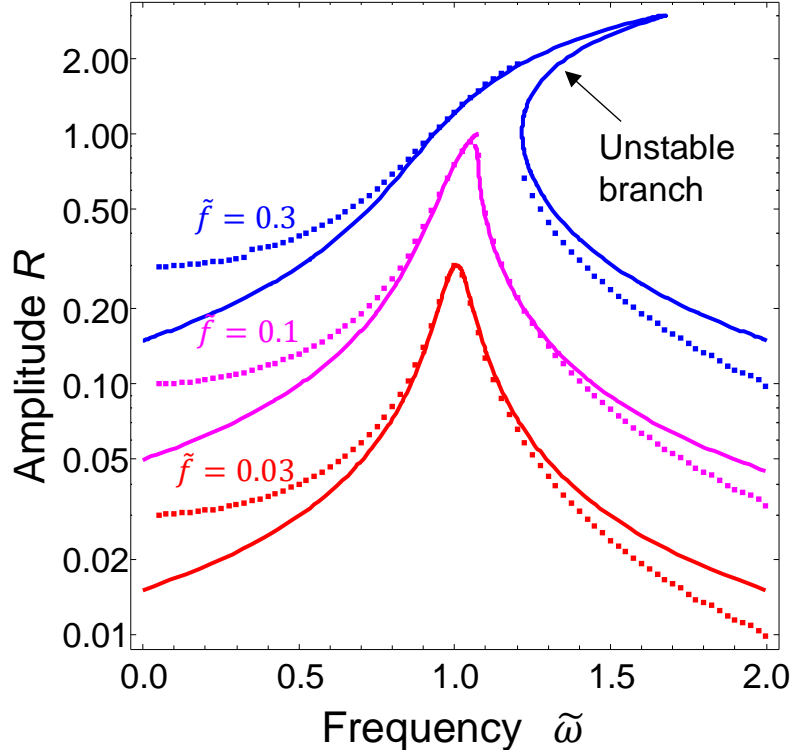

**Fig. S1.** Dependence of the oscillation amplitude on the driving force frequency for the different amplitude of force, $\tilde{f} = 0.03, 0.1, 0.3$ (red, magenta and blue curves respectively). Other parameters: $\tilde{c} = 0.1$, $\tilde{\lambda} = 0.2$. Dotted curves are calculated from the numerical solution of Eq. (A.2a), while solid curves are the graphical solution of Eq.(A.4a).

The equation (A.3) could be transformed into the dependence of amplitude R on frequency $\tilde{\omega}$. After some manipulations with (A.3) one could get the following:

$$R^2\left(4(\tilde{\omega}-1)^2 + \tilde{c}^2 + 3(1-\tilde{\omega})\tilde{\lambda}R^2 + \frac{9}{16}\tilde{\lambda}^2 R^4\right) = \tilde{f}^2 \qquad (A.4a)$$

Eq.(4a) represent bi-cubic equation with respect to amplitude $R$. Its evident solution is rather cumbersome; its six roots could be written as follows

$$R_1^2 = \frac{2\sqrt[3]{\tilde{D}}}{9\tilde{\lambda}^2} - \frac{8\left(3\tilde{c}^2 - 4(1-\tilde{\omega})^2\right)}{9\sqrt[3]{\tilde{D}}} + \frac{16(\tilde{\omega}-1)}{9\tilde{\lambda}} \qquad (A.4b)$$

$$R_2^2 = \frac{-(1-i\sqrt{3})\sqrt[3]{\tilde{D}}}{9\tilde{\lambda}^2} + \frac{4(1+i\sqrt{3})(3\tilde{c}^2 - 4(1-\tilde{\omega})^2)}{9\sqrt[3]{\tilde{D}}} + \frac{16(\tilde{\omega}-1)}{9\tilde{\lambda}} \qquad (A.4c)$$

$$R_3^2 = \frac{-(1+i\sqrt{3})\sqrt[3]{\tilde{D}}}{9\tilde{\lambda}^2} + \frac{4(1-i\sqrt{3})(3\tilde{c}^2 - 4(1-\tilde{\omega})^2)}{9\sqrt[3]{\tilde{D}}} + \frac{16(\tilde{\omega}-1)}{9\tilde{\lambda}} \qquad (A.4d)$$

Where the determinant was introduced as

$$D = \sqrt{\tilde{\lambda}^6(64(3\tilde{c}^2 - 4(\tilde{\omega}-1)^2)^3 + (-81\tilde{f}^2\tilde{\lambda} + 144\tilde{c}^2(\tilde{\omega}-1) + 64(\tilde{\omega}-1)^3)^2)} -$$

$$-16\tilde{\lambda}^3(9\tilde{c}^2 + 4(\tilde{\omega}-1)^2)(\tilde{\omega}-1) + 81\tilde{f}^2\tilde{\lambda}^4 \qquad (A.4e)$$

Let us try to drop higher order term $\sim \tilde{\lambda}^2 R^4$ in Eq.(A.4a):

$$R^2\big(4(\tilde{\omega}-1)^2 + \tilde{c}^2 + 3(1-\tilde{\omega})\tilde{\lambda}R^2\big) \approx \tilde{f}^2 \qquad (A.5a)$$

The solution of Eq.(5a) is

$$R^2 = \frac{1}{3(1-\tilde{\omega})\tilde{\lambda}}\left(-4(\tilde{\omega}-1)^2 - \tilde{c}^2 \pm \sqrt{(4(\tilde{\omega}-1)^2 + \tilde{c}^2)^2 + 3(1-\tilde{\omega})\tilde{\lambda}\,\tilde{f}^2}\right) =$$

$$= \frac{2\tilde{f}^2}{4(\tilde{\omega}-1)^2 + \tilde{c}^2 \pm \sqrt{(4(\tilde{\omega}-1)^2 + \tilde{c}^2)^2 + 3(1-\tilde{\omega})\tilde{\lambda}\,\tilde{f}^2}}$$

Only sign "+" matters:

$$R \approx \frac{\sqrt{2}\tilde{f}}{\sqrt{4(\tilde{\omega}-1)^2 + \tilde{c}^2 + \sqrt{(4(\tilde{\omega}-1)^2 + \tilde{c}^2)^2 + 3(1-\tilde{\omega})\tilde{\lambda}\,\tilde{f}^2}}} \qquad (A.5b)$$

The comparison of the "exact" graphical solution of Eq.(A.4a) with approximate solution (A.5b) is shown in **Fig. S2.**

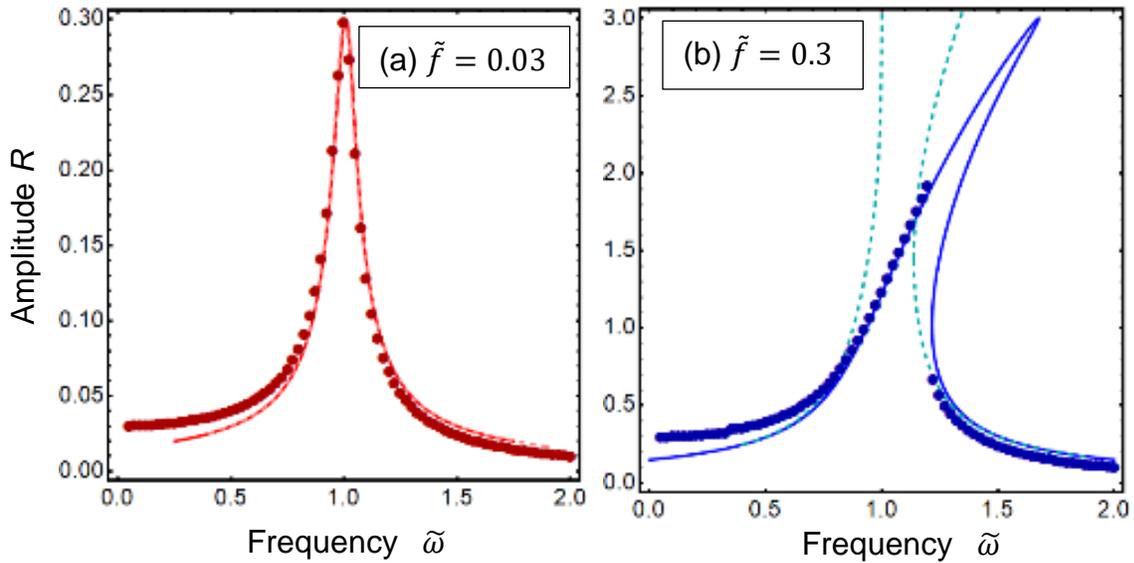

**Fig. S2.** Dependence of the oscillation amplitude on the driving force frequency for the different amplitude of force, $\tilde{f} = 0.03, 0.3$ (a, b). Other parameters: $\tilde{c} = 0.1$, $\tilde{\lambda} = 0.2$. Dots are for numerical solutions of Eq.(A.2a), solid curves are for Eq. (A.4), while dashed curves for Eq.(A.5b).

**Supplementary Table 1:**

Posterior estimates from SHO and Duffing models in Fig. 3(c,d). Recall that here we are fitting only to the Duffing cure in Fig. 3(a). The means and standard deviation are calculated from traces acquired during Metropolis sampling.

| | Parameter estimates (mean, standard deviation) |
|---|---|
| **Ground Truth Duffing (Fig.3a)** $[\tilde{f}, \tilde{c}, \tilde{\lambda}, \sigma^2]$ | [0.5, 0.2, 0.06, 0.15] |
| **Bayesian Duffing** $[\tilde{f}, \tilde{c}, \tilde{\lambda}, \sigma^2]$ | [(0.50595, 0.00771), (0.20368, 0.00478), (0.06187, 0.00447), (0.14577, 0.00774)] |
| **Bayesian SHO** $[f, \omega_r, Q, \sigma^2]$ | [(0.46963, 0.008122), (365660.62754, 786.43651), (5.66861, 0.15283), (0.15841, 0.00807)] |
| **Ground Truth Duffing (Fig.3b)** $[\tilde{f}, \tilde{c}, \tilde{\lambda}, \sigma^2]$ | [0.5, 0.2, 0.06, 0.90] |
| **Bayesian Duffing** $[\tilde{f}, \tilde{c}, \tilde{\lambda}, \sigma^2]$ | [(0.42377, 0.04416), (0.14595, 0.02393), (0.00250, 0.01701), (0.96926, 0.0.04952)] |
| **Bayesian SHO** $[f, \omega_r, Q, \sigma^2]$ | [(0.41994, 0.04706), (355608.53309, 3509.91011), (7.26482, 1.30975), (0.96385, 0.04975)] |

## Appendix B. Physical analogue of Duffing oscillator for BE-PFM

The linear partial differential equation defining the mechanical displacement vector $U$ for a (multi)ferroic film on a rigid substrate has the form:

$$\begin{cases} \frac{\partial \sigma_{ij}(x,t)}{\partial x_j} + \rho \frac{\partial^2 U_i}{\partial t^2}(x,t) = 0, \\ \sigma_{3i}(x_3 = 0) = 0, \quad U_i(x_3 = h) = 0. \end{cases} \quad (B.1)$$

Here $\rho$ is the mass density, $\sigma_{ij}$ is a stress tensor. In the limit of semi-infinite ferroic, the boundary condition $U_i(x_3 = h) = 0$ is substituted by the condition of stress absence. The time-dependent term $\rho \frac{\partial^2 U_i}{\partial t^2}$ is very small for excitation pulses at frequency $\omega$ much smaller than the characteristic frequencies of acoustic phonons, $\omega_a$, but otherwise it should be accounted for. For instance, it can be reasonable to account for this derivative for e.g. domain wall oscillations.

Assuming that the electromechanical coupling in a considered ferroic is described by an arbitrary tensor of piezoelectric and / or electrostrictive strains, generalized Hooke's law takes the form

$$\sigma_{ij} = c_{ijkl} u_{kl} - h_{ijk} P_k - q_{ijkl} P_k P_l, \quad (B.2a)$$

$$u_{ij} = s_{ijkl} \sigma_{kl} + d_{ijk} P_k + Q_{ijkl} P_k P_l, \quad (B.2b)$$

Where $c_{ijkl}$ and $s_{ijkl}$ are the tensors of elastic stiffness and compliances; $u_{kl} = \frac{1}{2}\left(\frac{\partial U_k}{\partial x_l} + \frac{\partial U_l}{\partial x_k}\right)$ is elastic strain tensor; $d_{ijk}$ and $h_{ijk}$ are the piezoelectric strain and stress tensors, that can be nonzero for a low symmetry paraelectric phase. Otherwise effective piezoelectric response appears in the ferroelectric phase only, as linearized electrostriction. $P_i(r,t)$ is a ferroic polarization. The electrostriction strain and stress tensors, $Q_{mjkl}$ and $q_{ijkl}$, can be renormalized by the Maxwell stresses.

LGD thermodynamic potential:

$$G_{LGD} = \int d^3r \left[ a P_i^2 + a_{ijkl} P_i P_j P_k P_l + a_{ijklmn} P_i P_j P_k P_l P_m P_n - E_i P_i - \frac{s_{ijkl}}{2} \sigma_{ij} \sigma_{kl} - Q_{ijkl} \sigma_{ij} P_k P_l - d_{ijk} P_k \sigma_{ij} + \frac{g_{ijkl}}{2} \frac{\partial P_i}{\partial x_j} \frac{\partial P_k}{\partial x_l} + \frac{F_{ijkl}}{2} \left( \sigma_{ij} \frac{\partial P_l}{\partial x_k} - P_l \frac{\partial \sigma_{ij}}{\partial x_k} \right) \right] \quad (B.3a)$$

Polarization components obey nonlinear dynamic equation of e.g. LGD type [ii]:

$$\mu \frac{\partial^2 P_i}{\partial t^2} + \Gamma \frac{\partial P_i}{\partial t} + 2aP_i - Q_{ijkl}\sigma_{kl}P_j + a_{ijkl}P_jP_kP_l + a_{ijklmn}P_jP_kP_lP_mP_n - g_{ijkl}\frac{\partial^2 P_j}{\partial x_k \partial x_l} =$$
$$d_{ijk}\sigma_{jk} + E_i. \quad \text{(B.3b)}$$

The coefficient $a(T)$ linearly depends on temperature $T$ and changes its sign at Curie temperature $T_C$. Kinetic coefficients $\mu$ and $\Gamma$ are responsible for the polarization stiffness (e.g. domain walls elasticity) and viscosity (Khalatnikov relaxation). The boundary conditions to Eq.(B.3a) are of the third kind and accounts for the flexoelectric effect:

$$\left(g_{ijkl}\frac{\partial P_k}{\partial x_l} - F_{klij}\sigma_{kl}\right)n_j\bigg|_{x_3=h} = 0 \quad \text{(B.3c)}$$

where **n** is the outer normal to the film surfaces, $F_{klij}$ is a flexoelectric effect tensor. Electric field $E_i(\mathbf{r},t)$ is the sum of external field, $E_i^{ext}(\mathbf{r},t)$, and internal depolarization field, $E_i^{dep}(\mathbf{r},t)$.

Equations (B.1) and (B.3) are coupled due to the generalized Hooke's law (2). Let us demonstrate the coupling using a "toy model" considering only the one-component ferroelectric polarization and elastic strain.

$$\frac{\partial \sigma}{\partial x} + \rho \frac{\partial^2 U}{\partial t^2} = 0, \quad \text{(B.4a)}$$

$$\sigma = cu - hP - qP^2 \cong \frac{1}{s}u - \frac{d}{s}P - \frac{Q}{s}P^2, \quad \text{(B.4b)}$$

$$\mu \frac{\partial^2 P}{\partial t^2} + \Gamma \frac{\partial P}{\partial t} + 2(a - Q\sigma)P + bP^3 + wP^5 - g\frac{\partial^2 P}{\partial x^2} = d\sigma + E(x,t), \quad \text{(B.4c)}$$

where $d$ is an effective piezoelectric constant, $Q$ is an electrostriction coefficient and $s$ is an elastic constant. For the considered one-component situation the relation between elastic, piezoelectric and electrostriction tensors are $c = \frac{1}{s}$, $h = \frac{d}{s}$ and $q = \frac{Q}{s}$. Let us consider uncharged domain walls, which do not induce local electric fields, so that $E(x,t) \to E_{ext}(t)$. For the case Eq.(B.4c) can be simplified as:

$$\mu \frac{\partial^2 P}{\partial t^2} + \Gamma \frac{\partial P}{\partial t} + 2\left(a - \frac{Q}{s}u\right)P + 3\frac{Qd}{s}P^2 + \left(b + 2\frac{Q^2}{s}\right)P^3 + wP^5 - g\frac{\partial^2 P}{\partial x^2} = \frac{d}{s}u + E_{ext}(t),$$
(B.5)

Next let us assume that we are in a deep ferroelectric phase, $P = P_S + \delta P$, where the spontaneous polarization $P_S$ is enough high, and so piezoelectric reaction dominates over electrostriction response, $\left|\frac{d}{s}P_S\right| \gg \left|\frac{Q}{s}P_S^2\right|$. Since the normal stresses are absent at the elastically free surface, the piezo-strain is $\frac{1}{s}u \approx \frac{d}{s}P + \frac{Q}{s}P^2 \approx \frac{d}{s}P$ in the region of BE-PFM response formation. Expressed via the piezo-strain Eq.(B.5) acquires the form

$$\mu \frac{\partial^2 u}{\partial t^2} + \Gamma \frac{\partial u}{\partial t} + \kappa(T)u + \gamma u^2 + \lambda u^3 + \chi u^5 - g\frac{\partial^2 u}{\partial x^2} = f_{ext}(t), \qquad (B.6)$$

Where $\kappa(T) = 2a(T) - \frac{d^2}{s}$, $\gamma = \frac{Q}{s}$, $\lambda = \left(b + 2\frac{Q^2}{s}\right)\frac{1}{d^2}$, $\chi = \frac{w}{d^4}$ and $f_{ext} = E_{ext}d$.

Eq.(B.6) is an equation in partial derivatives, it is much more complex than the ordinary differential Duffing equation (A.1), and so let us make several simplifying assumptions to establish correlations between them. Let us regard that the electrostriction contribution is negligibly small in comparison with effective piezoelectric coupling and soft mode nonlinearity, i.e. let us neglect $\gamma u^2$. Let us put $\chi = 0$ for the second order phase transition FE. Also, we assume that the spatial dispersion has a characteristic period $k_0$ in a Fourier space, and so we can estimate that $\frac{\partial^2 u}{\partial x^2} \approx -k_0^2 u$. After all these simplifications Eq.(B.6) acquires the form:

$$\mu \frac{\partial^2 u}{\partial t^2} + \Gamma \frac{\partial u}{\partial t} + ku + \lambda u^3 = f_{ext}(t), \qquad (B.7)$$

where $k(T) = 2a(T) - \frac{d^2}{s} + gk_0^2$, and $\lambda = \left(b + 2\frac{Q^2}{s}\right)\frac{1}{d^2}$. Note that Eq.(B.7) is a material equation, it can model the internal strain, but not the measured BE-PFM response, that is a convolution of the strain tensor with corresponding transfer function of a tip-surface junction.

Let us make some estimates of the coefficient $\lambda$ for the ferroelectric solid solution PbZr$_x$Ti$_{1-x}$O$_3$ (PZT, x<0.5), which can be the first order for x<0.2 and the second order for 0.3<x<0.5 at room temperature. Using its parameters from Refs.[iii, iv, v], we obtain the ranges for $Q \approx (0.01 - 0.05)$ m4/C2, $s \approx 1.5 \cdot 10^{-11}$ m2/N, $a \approx (10^8 - 10^9)$ m2N/C2, $b = \pm(0.05 - 5)10^7$m6N/C4, $g \approx 10^{-10}$C-2m4N, $k_0^2 \approx (1 - 10) \cdot 10^{18}$ m-2, and $d \approx (2.3 - 5) \cdot 10^{-2}$m2/C [vi], which give $k = \pm(0.5 - 5) \cdot 10^9$m2N/C2 and $\lambda = \pm(0.5 - 50) \cdot 10^9$ m2N/C2. Since the sum $\left(b + 2\frac{Q^2}{s}\right)$ can be negative, almost zero, or positive for the ferroelectrics with the first order phase transition in the bulk (i.e. when b<0), as well as change its sign due to the electrostriction coupling (adding positive term $2\frac{Q^2}{s}$), the dimensionless nonlinearity $\tilde{\lambda} = \left|\frac{\lambda}{k}\right| = (0.01 - 100)$ vary in a wide range, from realistically small to unrealistically high values.